\documentclass[12pt,letterpaper]{article} 
\usepackage{setspace}
\usepackage[top=1.7cm, left=2cm, bottom=2.5cm, right=2cm, paper=letterpaper]{geometry}
\usepackage{amsmath, amssymb, amsthm, amsfonts, graphicx}
\usepackage[utf8]{inputenc}
\usepackage[T1]{fontenc}
\usepackage{latexsym}
\usepackage{subfigure, float}
\usepackage{fancyhdr}
\usepackage{enumerate, enumerate, enumitem}
\usepackage[font={small}]{caption}
\usepackage{appendix}
\usepackage[nottoc, notlot, notlof]{tocbibind}
\usepackage{parskip}
\usepackage{setspace}
\usepackage{fontawesome}
\usepackage{booktabs}
\usepackage{makecell}
\usepackage{multirow}
\usepackage{pgfgantt}
\usepackage{tabularx}
\usepackage[document]{ragged2e}
\usepackage{color}
\usepackage{import}
\usepackage{lineno}
\usepackage[hidelinks]{hyperref}
\usepackage[none]{hyphenat}

\usepackage[
	authordate,
	backend=biber,
	url=false,
	doi=false,
	eprint=false,
	maxbibnames=10,
	maxcitenames=2,
	natbib
]{biblatex-chicago}
\usepackage{filecontents}

\DeclareFieldFormat[article,inbook,incollection,inproceedings,misc]{title}{#1\isdot}
\DeclareFieldFormat[article,inbook,incollection,inproceedings,misc]{number}{\addspace\bibstring{number}~#1\isdot}
\DeclareBibliographyDriver{article}{%
\printnames{author}%
\newunit\newblock
\printfield{year}%
\newunit\newblock
\printfield{title}%
\newunit\newblock
\printfield{journaltitle}%
\addspace
\printfield{volume}%
\newunit\addcomma
\printfield{number}%
\newunit\addcolon
\printfield{pages}%
\newunit\newblock
\printfield{note}%
\finentry}
\DeclareBibliographyDriver{misc}{%
\printnames{author}%
\newunit\newblock
\printfield{year}%
\newunit\newblock
\printfield{title}%
\newunit\newblock
\printfield{journaltitle}%
\addspace
\printfield{volume}%
\newunit\addcomma
\printfield{number}%
\newunit\addcolon
\printfield{pages}%
\newunit\newblock
\printfield{url}%
\newunit\newblock
\printfield{note}%
\finentry}

\hypersetup{
    colorlinks=false,
    linkcolor=black,
    urlcolor=black,
    citecolor=black
}

\newcolumntype{L}{>{\RaggedRight\arraybackslash}X}
\newcolumntype{Y}{>{\centering\arraybackslash}X}

\begin{document}

  %\linenumbers

    % Frontpage
    Methods Note/

    \Large{ \textbf{ Multiprocessing for the Particle Tracking Model MODPATH } } \normalsize
    Rodrigo Pérez-Illanes\\
    Corresponding author: Department of Civil and Environmental Engineering (DECA), Hydrogeology Group (UPC-CSIC), Universitat Politècnica de Catalunya, Jordi Girona 3-4, 08034, Barcelona, Spain, rodrigo.alfonso.perez@upc.edu 

    Daniel Fernàndez-Garcia\\
    Department of Civil and Environmental Engineering (DECA), Hydrogeology Group (UPC-CSIC), Universitat Politècnica de Catalunya, Jordi Girona 3-4, 08034, Barcelona, Spain, daniel.fernandez.g@upc.edu

    \textbf{Conflict of interest:} None
    \textbf{Article Impact Statement}: \textit{ Parallel particles processing for MODPATH using the OpenMP library}

    %% ABSTRACT
    \section*{Abstract}

Particle tracking has several important applications for solute transport studies in aquifer systems. Travel time distribution at observation points, particle coordinates in time and streamlines are some practical results providing information of expected transport patterns and interaction with boundary conditions. However, flow model complexity and simultaneous displacement of multiple particle groups leads to rapid increase of computational requirements. MODPATH is a particle tracking engine for MODFLOW models and source code displays potential for parallel processing of particles. This article addresses the implementation of this feature with the OpenMP library. Two synthetic aquifer applications are employed for performance tests on a desktop computer with increasing number of particles. Speed up analysis shows that dynamic thread scheduling is preferable for highly heterogeneous flows, providing processing adaptivity to the presence of slow particles. In simulations writing particles position in time, thread exclusive output files lead to higher speed up factors. Results show that above a threshold number of particles, simulation runtimes become independent of flow model grid complexity and are controlled by the large number of particles, then parallel processing reduces simulation runtimes for the particle tracking model MODPATH.

    %% INTRODUCTION
    \section*{Introduction}

MODPATH is a particle tracking post-processing program for MODFLOW-based groundwater flow models developed by the U.S. Geological Survey \citep{Pollock2016}. The program computes three-dimensional advective displacement of particles following the semi-analytical solution of \citet{Pollock1988}. Methodology allows particles to efficiently move towards a cell interface in a single displacement step, being then transferred to a connected neighbor cell. This process continues until particle encounters one of several possible stopping conditions, with displacement of one particle being independent of the others. 

Particle tracking has a variety of applications in studies of groundwater systems. Different simulation kinds provide necessary results for construction of travel time distributions (TTD) at observation points and for spatiotemporal characterization of particles and streamlines. These properties are often used to understand flow patterns in groundwater systems \citep[e.g.][]{Buxton1991}, delineate sources of water to discharging areas \citep[e.g.][]{Eberts2012}, determine time-dependent capture zones of wells \citep[e.g.][]{Bair1991,Riva2006}, and characterize the interplay between chemical reactions, dispersion and boundary conditions \citep[e.g.][]{Gusyev2014}, among others. Groundwater flow models with complex distribution of hydraulic properties or multiple boundary conditions might require simultaneous displacement of a large number of particles, and consequently, CPU demand. In reactive transport, different sets of particles are needed (one for each chemical substance), so computational requirements grow rapidly with increasing complexity of chemical systems. MODPATH source code \citep[see][]{Pollock2017} is written using serial programming, thus model runs are handled by a single CPU. Potential for parallelization has been identified in program stages processing particles due to independent displacements, which would allow to take advantage of all available computational resources simultaneously, leading to faster results. In this regard, a prototype GPU implementation of MODPATH algorithm has exemplified the potential for parallelization of this method \citep{Ji2020}.
 
Objective of this work is to incorporate distributed processing of particles into MODPATH starting from the current public version of the program \citep{Pollock2017}. Source code is written in Fortran and parallelization is implemented using the OpenMP library \citep{Openmp2020}. Performance of parallel implementation is compared against single processor runs for two synthetic test cases: a two dimensional heterogeneous aquifer with variable degree of heterogeneity and a three dimensional layered aquifer with multiple flow boundary conditions and different levels of grid complexity. Speed up is discussed for variable number of particles and processing threads, considering also different OpenMP library configuration scenarios. 

The paper is organized as follows. Methods section presents a discussion of MODPATH flow chart, followed by a summary of considerations regarding possible OpenMP configurations and presentation of synthetic test cases. Results and discussions elaborate from the speed up quantification of endpoint and timeseries simulations of test cases. Conclusions summarize results from implementation, revisiting interaction between aquifer model characteristics and parallel library configuration.

    %% METHODS
    \section*{Methods}

\subsection*{MODPATH}
Current version of the software is implemented in Fortran following an object oriented programming paradigm and works as a module independent of MODFLOW \citep{Pollock2016}, thus the particle tracking process is decoupled from the groundwater flow model. It provides compatibility for models based on MODFLOW-2005 \citep{Harbaugh2005} and MODFLOW-6 \citep{Langevin2017}, for both structured and rectangular unstructured grids \citep[see][]{Pollock2015}.

MODPATH displaces particles through flow model cells until an stopping condition is met. These include for example encountering a boundary face, reaching the maximum tracking time, or landing in a cell with sink flows. The latter depends on the specific configuration of the particle tracking model because in some scenarios, mostly while considering flow models with coarse resolution, particles might be still allowed to displace inside these cells, in which case these are known as weak sinks. Different approaches for modeling weak sink cells have been addressed in literature \citep{Visser2009, Abrams2012}. Since the displacement of one individual particle is independent of the others, there is significant potential for parallelization, in particular, for particle tracking models with high computation requirements. For example, a transient groundwater flow model of an aquifer with a complex distribution of hydraulic properties, multiple flow boundary conditions and simultaneous displacement of several particle groups, is an scenario easily found in reactive transport analysis of hydrogeologic units.

% FIGURE 1
\begin{figure}[ht!]
    \centering
    \includegraphics[scale=1]{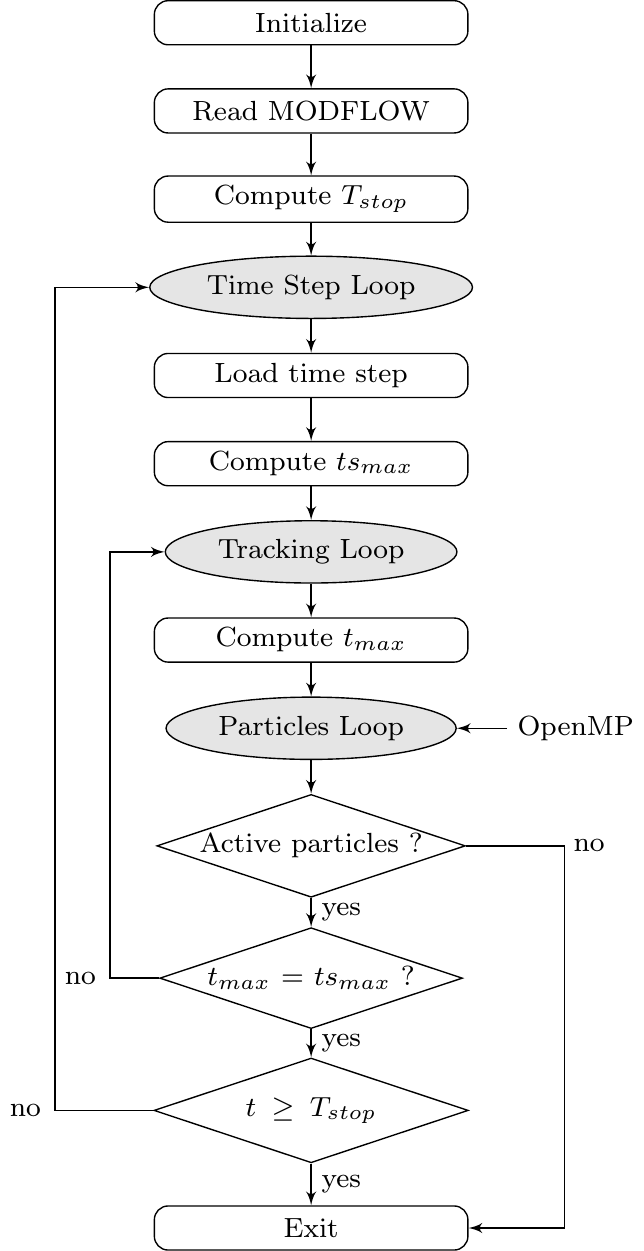}
    \caption{MODPATH simplified flow chart. Particles loop is parallelized with OpenMP.}
    \label{fig:flowchart}
\end{figure}

Program initializes reading a configuration file that specifies data related to MODFLOW model and particle tracking process (Fig. \ref{fig:flowchart}). Depending on the type of simulation (endpoint, timeseries, pathline) and configuration parameters, a simulation stop time $T_{stop}$ is determined. The program has three relevant nested loop structures. The outermost structure is the time step loop, which runs over the MODFLOW time steps individually, defined from all of the flow stress periods. For each loop, MODPATH computes the maximum time for the current flow conditions $ts_{max}$ and once this limit is reached, flow related arrays are updated. The middle loop structure is the tracking loop, which sets the maximum tracking time $t_{max}$, depending on the simulation type. This time is not necessarily the same as $ts_{max}$. For example, in a timeseries simulation it is possible that the user defines smaller output time steps than the time step of the flow model. In this case, the program will perform several tracking stages until $ts_{max}$ is reached. The innermost loop runs over particles, where each one is displaced individually and this is the stage with highest potential for parallelization. At this point, it is important to remark that algorithm for particles displacement and determination of stopping conditions while considering the parallel implementation, remains exactly the same as the original MODPATH. For more details about these procedures the reader is referred to \citet{Pollock1988,Pollock2016}. 

In MODPATH, timeseries simulations can be chosen to visualize the temporal evolution of particle clouds. This is often used to understand, for instance, the interaction between transport and flow patterns in groundwater systems. Usually, these type of simulations are configured in such a way that particles positions are written to output files at user defined times or equispaced by a given time interval. These simulations are particularly challenging when considering multiprocessing because the writing to output files is performed within the particles loop. This and other related aspects when considering multiprocessing for the program are discussed in the following.

\subsection*{Multiprocessing}
Practical considerations have arisen from integrating multiprocessing into MODPATH due to the interactions between the program structure and the OpenMP library configurations. Before discussing these in detail, some concepts related to OpenMP need to be introduced. Specifically, sections of the code executed in parallel are known as \emph{parallel regions} and a single process within these sections is known as a \emph{thread} \citep{Openmp2020}. Throughout this article, multiprocessing and parallel processing are used indistinctively to indicate that tasks are distributed among several threads on a single computer, with the concept of threads closely related to computing cores.

In MODPATH, particles are displaced by a tracking engine represented in source code by an object class that manages the displacement procedure between flow cells and verifies stopping conditions. Current implementation of tracking engine contains flow related arrays as class properties used for the initialization of flow cell velocities. In order to parallelize MODPATH, the tracking engine needs to be independent for each processing thread to avoid memory inconsistencies. That is, each thread has to manage a different particle history and current flow model cell. OpenMP library allows defining thread private objects by creating independent copies managed by each thread. Replication of the tracking engine in its current form would also mean that flow related arrays are replicated unnecessarily inside the parallel region. Since flow models may be composed of millions of cells, this can be memory demanding. To overcome this problem, the implementation of particles multiprocessing considers the introduction of an intermediate object that handles flow model information, which is stored centrally. The latter is accessed from the tracking engine through a pointer which can be easily replicated without significant impacts on system memory.  

Similarly, some considerations should be taken into account for timeseries and pathline simulations. For these, MODPATH writes output records while being inside the particles loop. As each processing thread operates independently, it is possible that more than one thread tries to write a record to an output unit, which may generate data corruption if not handled appropriately. For such situations, OpenMP library provides thread exclusive clauses that block a portion of code execution for other threads if one is performing such instructions. Nevertheless, when thread exclusive operations are performed with high frequency, blocking clauses may reduce the efficiency of parallel implementation. For comparison purposes, three different output procedures are discussed for timeseries simulations: (i) all threads write exclusively to a single output unit using OpenMP \texttt{critical} directive; (ii) threads write to specific binary output units which are then consolidated into a text-plain file at the end of timeseries step; and (iii) thread exclusive text-plain output units are not consolidated. In the case of the consolidated protocol, the output of thread specific units is recollected after each timeseries step in order to preserve sorting of time indexes. 

Another important aspect to consider for multiprocessing is the workload distribution protocol while MODPATH is executing the parallel loop. This is known as \textit{scheduling} and specifies the distribution of loop indexes (particles) to be processed by each thread. In this article, two scheduling strategies are discussed. The first one is the \texttt{static} scheduling with balanced distribution of particles. In this case, the number of iterations is approximately the same for each thread and particles to be processed are specified only once before entering the parallel region. It is a good approach for models where simulation time is approximately the same for all particles. The latter however, is not necessarily true in real groundwater systems.

Particles are displaced sequentially between flow model cells, involving a cell initialization stage before computing displacements. This means that the computational time required for processing the trajectory of particles can be different, most notably, in scenarios with non-uniform flow where the length of streamlines is influenced by the spatial variability of hydraulic properties. To exemplify this, consider a timeseries simulation writing particles position at regular time intervals, using a model aquifer with non-uniform velocity distribution. A particle moving along a streamline of relatively high velocity will travel through a higher number of cells compared to slower particles for the same simulation time interval. Consequently, faster particles will require the initialization of a higher number of flow cells. Differences in loop time for each particle may be further influenced by outflow boundaries that also act as stop conditions.

Different processing times for particles motivate the analysis of \texttt{dynamic} scheduling, which distributes particles to be processed by each thread during runtime. A new particle is assigned after the completion of the current. The advantage of this approach is that it reduces the likelihood of threads being idle. A thread processing a slow particle, at the end of the simulation, might have processed a smaller total amount of particles than others, without necessarily delaying the total simulation time. The possibility of unbalanced distribution of processed particles, gives to the program some adaptivity to the particles travel time distribution. This can favor some hydrogeological settings, like for example a highly heterogeneous system. The distribution of particles during runtime in \texttt{dynamic} scheduling introduces some overhead in comparison to \texttt{static} scheduling. The impact that this effect might have in MODPATH simulation times is discussed in the following sections.

\begin{figure}
    \centering
    \includegraphics[scale=1]{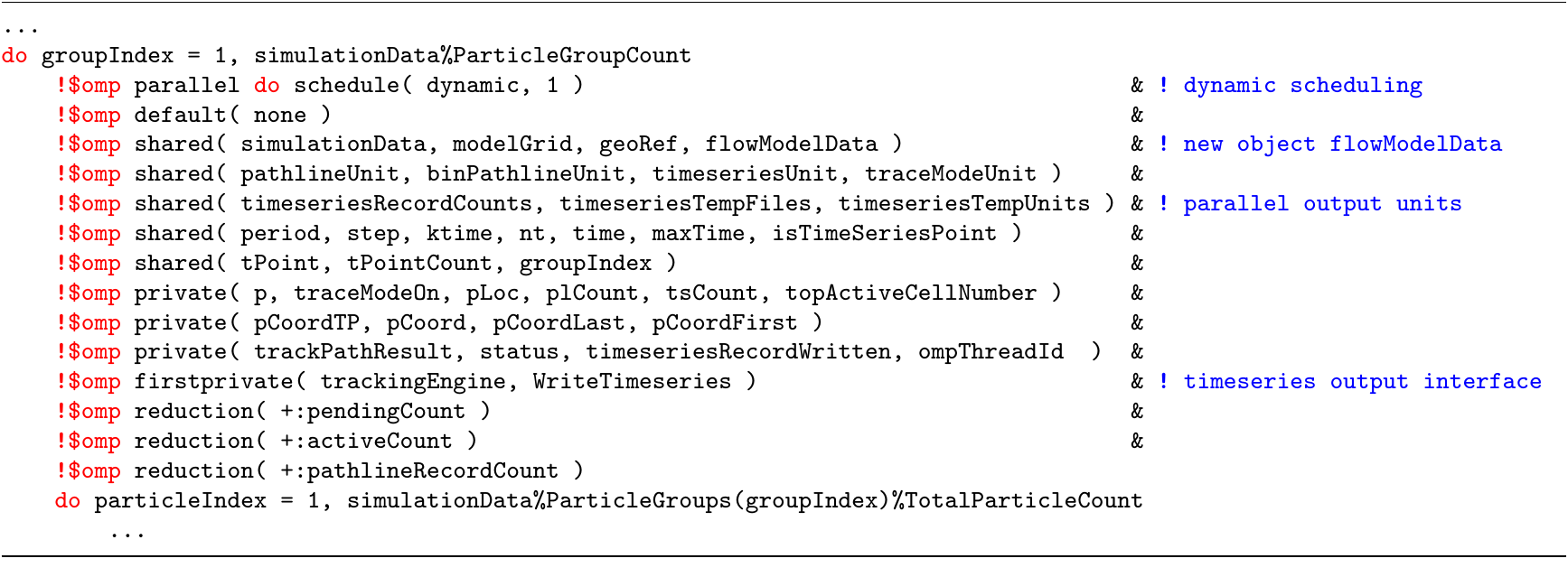}
    \caption{OpenMP specification for parallel particles loop in MODPATH. Memory state for all loop variables is explicitly declared as \texttt{shared}, \texttt{private} or \texttt{firstprivate}. Counters are declared with a summation \texttt{reduction} clause.}
    \label{fig:openmp}
\end{figure}

An additional aspect to consider for multiprocessing is that MODPATH stores the particles in a list, assigning particles' indexes according to the position in which they appear in this list. In the serial implementation particles are processed sequentially, which means that output files will present sorted particles indexes. With the parallel implementation the list of particles is not necessarily processed in order, hence output files will not display sorted particles indexes regardless of the scheduling protocol. Because particles are displaced independently, this does not cause issues in the context of the program, however leads to visible differences in output files, so attention should be paid in cases where post-processing tools rely on sorted particles indexes while reading output files from parallel model runs. At this point, it is important to remark that output results from single thread and parallel runs of models discussed in the following sections were verified to be equivalent in value. For reference, OpenMP specification enabling parallel particles loop is shown in Figure (\ref{fig:openmp}). By default OpenMP library considers that all variables within parallel loops are \texttt{shared} between threads, except the loop index. However, in some cases this may lead to memory inconsistencies specially in complex loop structures. To avoid this issue, memory state of all loop variables is forced to be explicitly defined with the clause \texttt{default(none)}. Depending on their functionality, memory states are declared as \texttt{shared}, thread \texttt{private} or initialized with the same values for all threads and then private (\texttt{firstprivate}). Similarly, counters increasing their values within the parallel loop should be declared with a \texttt{reduction} clause for consistent results.

% FIGURE 2

\subsection*{Synthetic test cases}
This section presents synthetic test cases aimed at evaluating the performance of MODPATH parallelization under different scenarios. Simulations are performed on a desktop computer with an Intel\textsuperscript{\textregistered{}} Core\texttrademark{} i7-9700 CPU @ 3.00GHz processor. For each test case, the number of processing threads $N$ is modified in integer powers of 2 between $1-8$ and parallel MODPATH code is compiled with \texttt{gfortran@9.2.1} \citep{Gfortran2020} on a \texttt{linux} system. CPU performance is compared considering the elapsed time reported by MODPATH, which measures the time employed by the outermost loop (Time Step Loop in Fig. \ref{fig:flowchart}), including the writing to output files in the case of timeseries and pathline runs, but not the reading of input configuration and MODFLOW files. In this regard, an additional variable that could influence overall performance is the kind of disk where data is being written (Solid State Drive, SSD; or Hard Disk Drive, HDD). This would not be the case however for the time reported in endpoint simulations, because output files in this case are written after the outermost loop. For the simulations of this article, observed speed up due to parallelization was in general similar while writing to SSD or HDD, with the exception of one scenario addressed with more detail in results and discussion.

\begin{figure}[]
    \centering
    \includegraphics[scale=1]{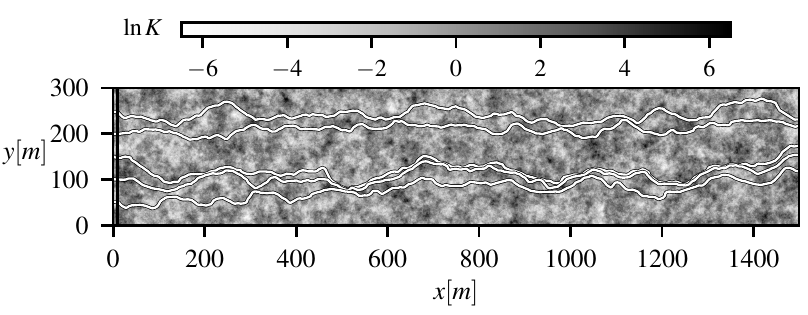}
    \caption{Synthetic two-dimensional heterogeneous aquifer (TC1). Vertical black line near the origin marks particles injection and white lines show reference streamlines for $\sigma^2_Y=2.5$.}
    \label{fig:streamlines}
\end{figure}

The first test case (TC1) consists of a two-dimensional heterogeneous aquifer under steady-state flow conditions (Fig. \ref{fig:streamlines}). Objective of this test is to evaluate performance of parallel MODPATH under different scenarios of heterogeneity, and hence, spatial variability of flow velocities. For these purposes, a domain discretized in $1500\times300$ cells of size $\Delta_x=\Delta_y=1[m]$ is considered (Table \ref{tab:tc1}), with a spatial distribution of hydraulic conductivity constructed from one realization of a sequential Gaussian simulation, denoted as $Y(\bold{x})$, characterized by an isotropic exponential variogram with correlation length of $I_Y=10[m]$, zero mean, and unit variance. The hydraulic conductivity is determined according to $K(\bold{x}) = \exp\left(\sigma_{Y} Y(\bold{x}) \right)$. The degree of heterogeneity is therefore controlled by the variance $\sigma^2_{Y}$, while preserving the underlying conductivity patterns across test runs. Groundwater flow is induced by a unit mean pressure gradient oriented along the $x$-axis. Models are solved with MODFLOW-6 and convergence is verified for each simulation. Particles are injected near the aquifer inlet, uniformly distributed at $x_o = 10[m]$. The number of injected particles $N_p$ is increased between $10^3-10^7$. A set of endpoint simulations for different values of $\sigma^2_Y$ are used to compare thread scheduling strategies. Simulations are configured to displace all particles until reaching the aquifer outlet.

% FIGURE 3

\begin{table}[ht!]
  \centering
  \caption{Parameters for synthetic test case TC1.}
  {\small
    \begin{tabularx}{8.2cm}{@{} l l c*{2}{Y}}
      \toprule
             \multicolumn{2}{l}{Parameter}                 & Value             & Unit  \\  
      \midrule  
	Cell size                     & $\Delta_{x,y}$     & $1$               & $m$     \\
        Correlation length            & $I_Y$              & $10$              & $m$     \\
        Aquifer length                & $L_x$              & $1500$            & $m$     \\
        Aquifer width                 & $L_y$              & $300$             & $m$     \\
        Inlet constant head           & $C_h^{in}$         & 1510              & $m$     \\
        Outlet constant head          & $C_h^{out}$        & 10                & $m$     \\
        Aquifer variance              & $\sigma_Y^2$       & $0.1 - 5$         & -     \\
      \bottomrule                                                       
    \end{tabularx}
  }
  \label{tab:tc1}
\end{table}

\begin{figure}[]
    \centering
    \includegraphics[scale=1]{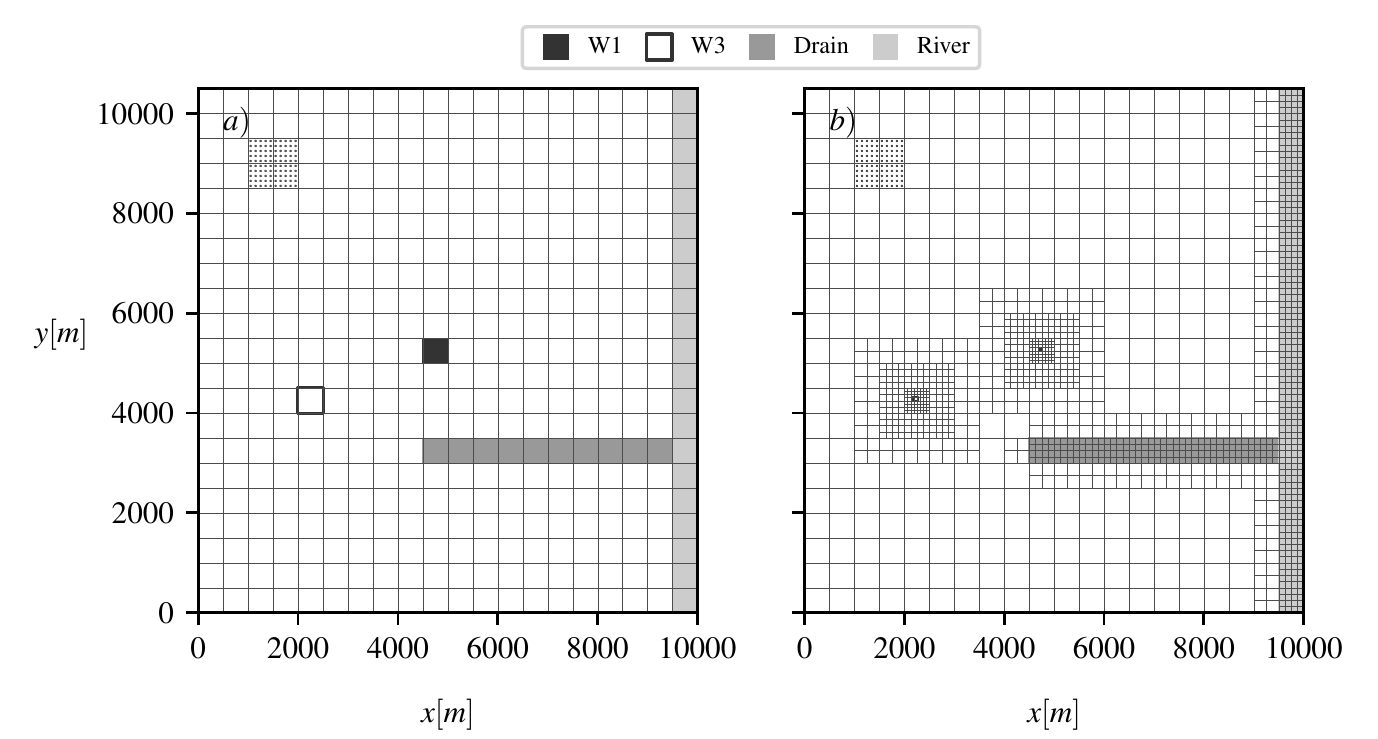} 
    \caption{Synthetic three-dimensional layered aquifer (TC2). $a)$ Original structured grid, $b)$ modified unstructured grid. In both panels, scatter points indicate the particles release area.}
    \label{fig:complexdiagram}
\end{figure}

The second synthetic test case (TC2), considers a layered three-dimensional aquifer subject to several flow boundary conditions. The objective of this test case is to analyze the improvement due to parallelization in more practical and complete groundwater model scenarios, involving recharge, pumping wells, river and drain boundary conditions. The problem is based on one of the examples provided in the current MODPATH repository \citep[{\texttt{ex03\_mf6} in}][]{Pollock2017}, with model parameters reinterpreted in SI units (Table \ref{tab:tc2}). The example has been modified with grid refinement near relevant boundary conditions to also consider an unstructured \textit{quadtree} grid (Fig. \ref{fig:complexdiagram}). The domain is composed of three homogeneous layers that represent two aquifers separated by an aquitard. Hydraulic conductivity for both top and bottom aquifers present vertical anisotropy ($K_{zz}<K_{xx},K_{yy}$), while the two horizontal values are considered to be the same. The original structured grid was composed of $N_{cpl}^{str}=420$ cells per layer, but after refinement this number has grown to $N_{cpl}^{usg}=1464$. The system is subjected to homogeneous recharge with an east-side river boundary condition. The problem analyzes trajectories of particles released from the surface layer, influenced by two pumping wells acting individually on the top and bottom aquifer (W1 and W3 respectively). The default test case configuration allows particles to pass through weak sink cells. MODPATH simulation considers timeseries runs with 5 and 30 time snapshots of particles location (TS5 and TS30 respectively) to quantify performance improvements under different output conditions. Snapshots are homogeneously distributed along the timeseries simulation time. Unstructured grid simulations are also performed considering the same output conditions. Like previous test case, the total number of injected particles is systematically increased. Particles are uniformly released from the uppermost face of four cells in the top aquifer, injected in 10 stages every 20 days of simulation time. The final injection time is small compared to the final timeseries time ($T_{ts}=60000 [d]$, $T_{inj}/T_{ts}\approx0.3\%$). This test considers 3 flow stress periods, although values for boundary conditions remain the same during the simulation: an initial steady-state, a second transient flow with 10 time steps, and a final stress period also at steady-state. This means that for TC2, MODPATH will employ some simulation time in updating the flow model arrays, which is currently performed in serial. This reduces the potential speed up factor due to parallelization of the particles loop because in the total elapsed time used for comparison, there is a higher proportion of mandatory serial operations in comparison to a fully steady-state flow model.

% FIGURE 4

\begin{table}[ht!]
  \centering
  \caption{Parameters for synthetic test case TC2. Layered properties are shown in curly brackets from top to bottom layer.}
  {\small
    \begin{tabularx}{8.2cm}{@{} l l c*{2}{Y}}
      \toprule
             \multicolumn{2}{l}{Parameter}          & Value              & Unit             \\  
      \midrule                                                                              
    	Flow rate W1       &  $Q_{W1}$          & -$7.5\times10^4$   & $m^3/d$           \\ 
    	Flow rate W3       &  $Q_{W3}$          & -$1\times10^5$     & $m^3/d$           \\
    	Hydr. cond.        &  $K_{xx,yy}$       & \{50 , 0.01, 200\} & $m/d$             \\
    	Hydr. cond.        &  $K_{zz}$          & \{10 , 0.01, 20 \} & $m/d$             \\
            Layer height       &  $\Delta_z$        & \{130, 20  , 200\} & $m$               \\
            Spec. yield        &  $S_y$             & 0.1                & -                 \\
    	Spec. storage      &  $S_s$             & $1\times10^{-4}$   & $1/m$             \\
    	Recharge           &  $q_R$             & $5\times10^{-3}$   & $m/d$             \\
            River stage        &  $R_s$             & 320                & $m$               \\
            River bot.         &  $R_b$             & 317                & $m$               \\
    	River cond.        &  $C_R$             & $1\times10^{5}$    & $m^2/d$           \\
            Drain elev.        &  $D_e$             & 322.5              & $m$               \\
    	Drain cond.        &  $C_D$             & $1\times10^{5}$    & $m^2/d$           \\
      \bottomrule                                                       
    \end{tabularx}
  }
  \label{tab:tc2}
\end{table}

    %% RESULTS AND DISCUSSION
    \section*{Results and discussion}

\subsection*{Thread scheduling}
\begin{figure}[]
    \centering
    \includegraphics[scale=1]{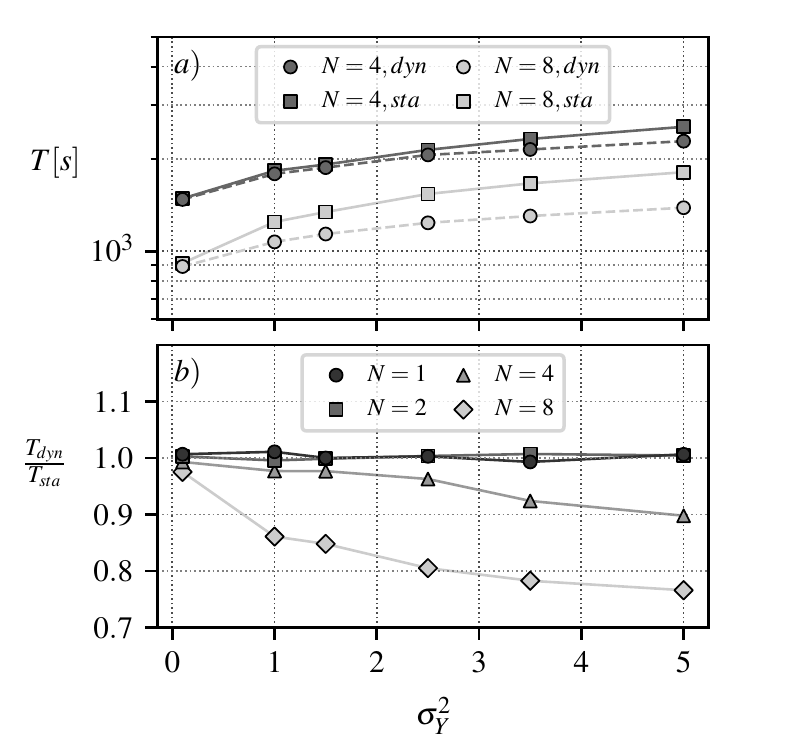}
    \caption{Performance of endpoint simulations with $N_p=10^7$ particles as a function of aquifer variability and different number of threads $N$. $a)$ Computational time for \texttt{static} (solid, squares) and \texttt{dynamic} (dashed, circles) scheduling with 4 and 8 threads; $b)$ measured ratio $T_{dyn}/T_{sta}$ for all thread configurations. }
    \label{fig:omp:endpoint}
\end{figure}

Runtimes from TC1 simulations obtained with different thread scheduling strategies are used to determine the best protocol for parallelizing MODPATH under different scenarios of heterogeneity. Regions of low and high pixel values of the aquifer realization $Y(\bold{x})$ will have lower and larger hydraulic conductivity values, respectively, when increasing the aquifer variance $\sigma_Y^2$. For a fixed pressure gradient, this means that minimum and maximum flow velocities are also influenced by this parameter. MODPATH will displace all particles until reaching the aquifer outlet. So, increasing the variance leads to longer simulation times, as shown in Figure (\ref{fig:omp:endpoint}$a$). This occurs consistently for both \texttt{static} and \texttt{dynamic} thread scheduling protocols. Simulation times with the latter are lower for all scenarios of heterogeneity, indicating that improvements in time due to dynamic scheduling are able to compensate the expected overhead from the assignment of particles during runtime.

Increasing the number of processing threads leads to significant differences in the runtime obtained with different threading protocol. The ratio between simulation times $T_{dyn}/T_{sta}$ for a fixed number of particles is shown in Figure (\ref{fig:omp:endpoint}$b$). In runs with 1 or 2 threads, the ratio is close to 1, and the impact of scheduling strategy is not noticeable, even for large aquifer variances. However, simulations with 4 and 8 threads exhibit an important reduction of the computational time when \texttt{dynamic} scheduling is used. For higher values of aquifer variability and $N=8$, the simulated ratio indicates that \texttt{dynamic} simulations require up to $25\%$ less runtime than the corresponding \texttt{static} simulation for the variances investigated, illustrating the interplay between the threading protocol (a parallel library configuration) and aquifer heterogeneity on the total simulation time.

% FIGURE 5

\begin{figure}[]
    \centering
    \includegraphics[scale=1]{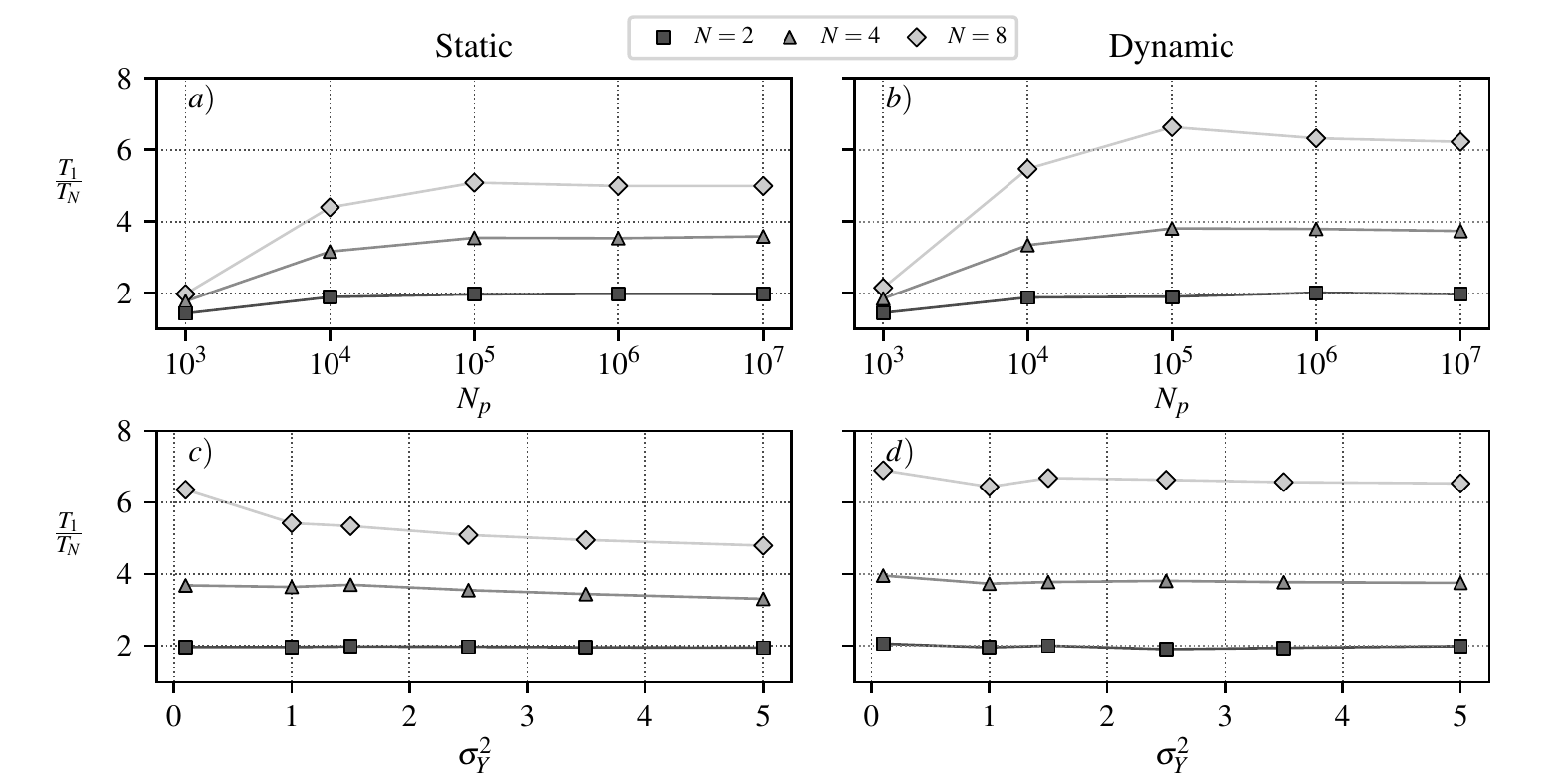}
    \caption{Speed up for each threading protocol. $a),b)$ as a function of the number of particles with aquifer variance of $\sigma_Y^2=2.5$, $c),d)$ as a function of aquifer variance with number of particles $N_p=10^7$.}
    \label{fig:omp:endpoint:speedup}
\end{figure}

The speed up factor relating simulation times between runs with single thread and parallel simulations $T_1/T_N$ provides a complementary picture of the parallelization performance. Results show (Fig. \ref{fig:omp:endpoint:speedup}$a,b$) that for a given aquifer variance, speed up factor is limited by the number of particles. Above $N_p=10^5$, the speed up approaches an asymptotic value, which occurs for both scheduling protocols. However, the speed up displays higher values when employing \texttt{dynamic} scheduling for simulations with number of particles above $N_p=10^5$. For this number of particles, $\sigma_{Y}^2=2.5$, and maximum number of threads, the speed up of the \texttt{static} and \texttt{dynamic} scheduling protocols were $T_1/T_8=5.09$ and $T_1/T_8=6.63$, respectively. Notice that expected speed up values from this implementation of parallel MODPATH, will not be as high as those observed in prototypical GPU codes \citep[{e.g.}][]{Ji2020}, where the hardware can provide thousands of processing threads, in contrast to classical desktop CPU hardware. Still, the development here presented comes with the inherent advantage that is integrated into MODPATH source code, without any change to current input files, then adoption is straightforward.

Another interesting aspect of the speed up factor and the threading protocol is its dependence on the degree of heterogeneity $\sigma_Y^2$ (Fig. \ref{fig:omp:endpoint:speedup}$c,d$). When using \texttt{static} scheduling and maximum threads, the speed up factor decreases consistently with the degree of heterogeneity, which is explained by both larger travel times to aquifer outlet and the even distribution of particles to be processed. Under this scenario of workload distribution, consider a simplified case of only one slow particle and all the rest moving at the same velocity, then it is evident that threads will be waiting for completion of the slow particle. In contrast, speed up factors achieved with \texttt{dynamic} scheduling are essentially independent of aquifer variance for a given number of particles in test case TC1.

% FIGURE 6

\subsection*{Timeseries}
Performance of endpoint simulations provided a base line for discussion of timeseries, given that in some cases, endpoint runs can be seen as timeseries with no output stage. Still, printing out particle positions on runtime requires some considerations in order to preserve parallel speed ups seen in the corresponding endpoint simulations.

\begin{figure}[ht!]
    \centering
    \includegraphics[scale=1]{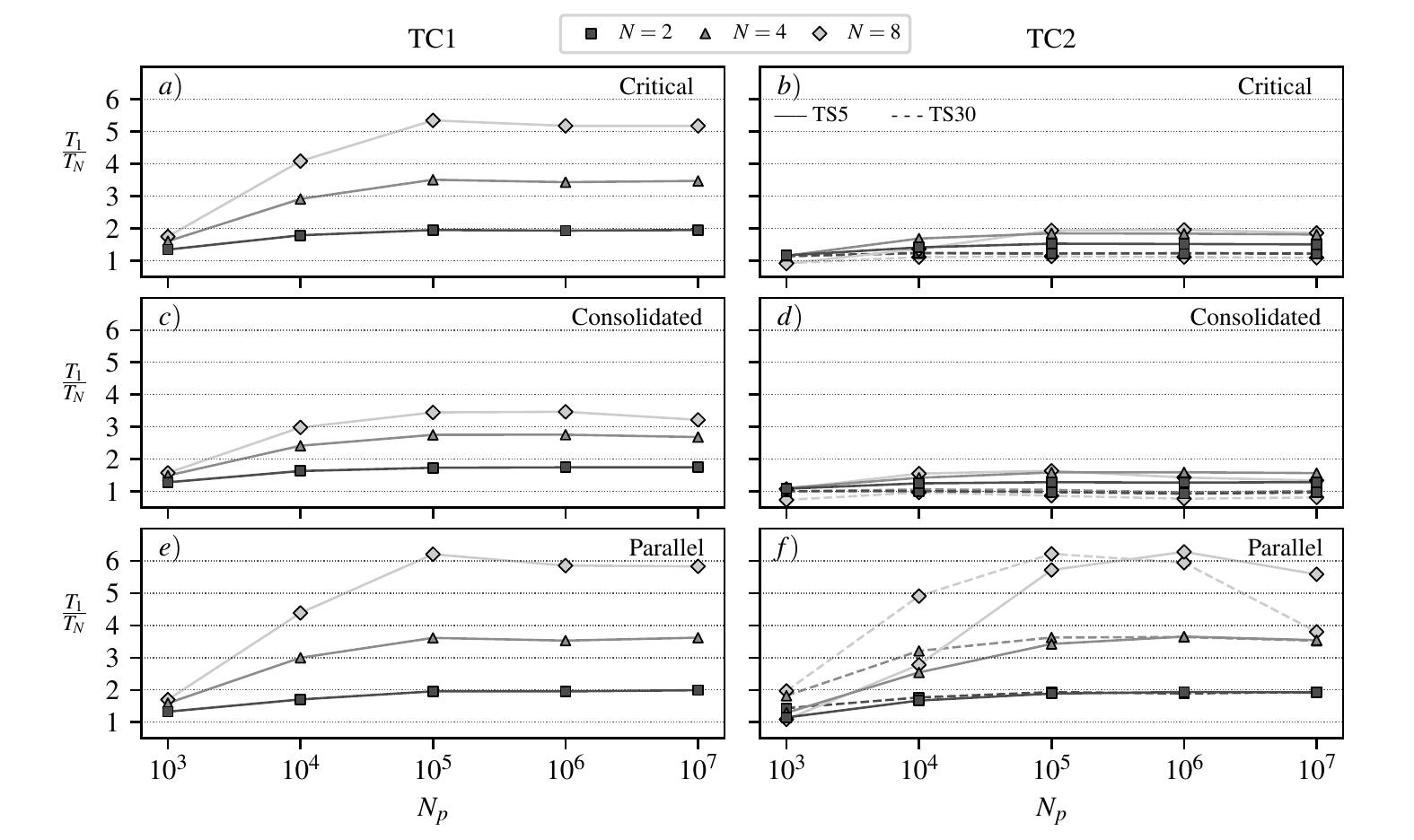}
    \caption{Speed up factors for parallel output protocols. Panels $a),c),e)$ (first column) present results for timeseries in test case TC1 with 10 writing stages considering $\sigma^2_Y=2.5$. Panels $b),d),f)$ (second column) display results for test case TC2 with 5 (solid, TS5) and 30 (dashed, TS30) writing stages. Corresponding output protocol is indicated in each panel.}
    \label{fig:tsout}
\end{figure}

The obtained speed up factors are different depending on the output protocol. To exemplify this, a timeseries simulation with \texttt{dynamic} scheduling is performed using TC1 considering $\sigma_Y^2=2.5$ and 10 writing stages. Results show (Fig. \ref{fig:tsout}$a,c,e$) that parallel output leads to higher values of speed up in comparison to the other output protocols, with magnitudes close to those obtained in endpoint runs. Employing the blocking clause (\texttt{critical}) leads to smaller speed up factors, however, without much impact in comparison to the case of parallel output files. In constrast, the consolidated protocol is the slowest due to the time required to load thread specific data and write it back to the consolidated file after each timeseries snapshot.

% FIGURE 7

From timeseries runs using test case TC2 with unstructured grid, considering 5 and 30 writing stages, different performance is observed (Fig. \ref{fig:tsout}$b,d,f$). Specifically, employing critical or consolidated output formats leads to a significant decrease in the speed up factor relative to parallel output. Moreover, in some cases of consolidated output protocol and TS30, the performance decreases in comparison to single thread runs. Parallel output preserves the magnitude of speed up observed in previous numerical tests. Notably for this case, changes in speed up factor with respect to the number of particles exhibit a peak value when employing the maximum number of threads, with the number of particles of maximum performance depending on the timeseries output frequency. Smaller frequency (TS5) leads to maximum speed up factor with a higher number of particles ($N_p=10^6$) than the obtained for the simulation with high frequency (TS30). In this case the number of particles of maximum speed up is $N_p=10^5$. This effect could be explained by differences on disk writing speeds. Simulations with high output frequency and high number of particles are expected to generate large output volumes. When writing in parallel, all threads write large volumes of data simultaneously generating simulation output faster than single thread runs. Simulations with parallel output protocol shown in Figure (\ref{fig:tsout}$f$) were performed writing to HDD hardware, and the decreasing speed up for high number of particles may be explained by limitations on output writing speed. For comparison, TC2 runs with the highest output frequency (TS30) and parallel output protocol were also configured to write to output files in SSD hardware. Speed up factors where similar to those obtained from HDD, excepting the case with highest number of particles ($N_p=10^{7}$) and $N=8$ threads. In this scenario, speed up when writing to HDD was $T_1/T_8=3.8$ (Fig. \ref{fig:tsout}$f$), whereas when writing to SSD reached a value of $T_1/T_8=5.1$. A similar analysis was performed for the other two output protocols in TC2, with small influence of disk hardware on speed up results, although in general faster simulations with SSD.

There are significant differences in timeseries performance between the two test cases while employing the blocking directive for output files. Besides the time required in TC2 for updating flow model arrays due to the transient stress period, this difference in speed up is explained by the influence that heterogeneity in TC1 has on the simulation time required for a particle to reach the timeseries output time. As seen previously, non-uniform flow leads to differences in the processing time for each particle, depending on the characteristics of the streamline and influenced by the number of cells initialized during displacement. This means that particles traveling through streamlines with different velocities, will arrive to the instant where the output is required at different times. This decreases the likelihood of the output file being busy because another thread is writing. To remark this point, the opposite case of a fully homogeneous domain can be considered, which is closer to the conditions in TC2. The processing time needed to reach the timeseries writing stage for all particles (traveling in horizontal streamlines with the same velocity) will be always almost the same. As a result, all threads will try to write simultaneously to the output unit, and the blocking clause will force some threads to wait. 

In any case, results from both test cases show that parallel output with thread exclusive units is by far the fastest approach and provides significant speed ups for increasing number of threads and particles. Thus, in general, writing to thread specific output units reduces the simulation runtime, however, this format requires reading thread specific output files during post-processing stages. These files have been configured to preserve the same data structure used in the original MODPATH program.

\subsection*{Grid complexity}
\begin{figure}[ht!]
    \centering
    \includegraphics[scale=1]{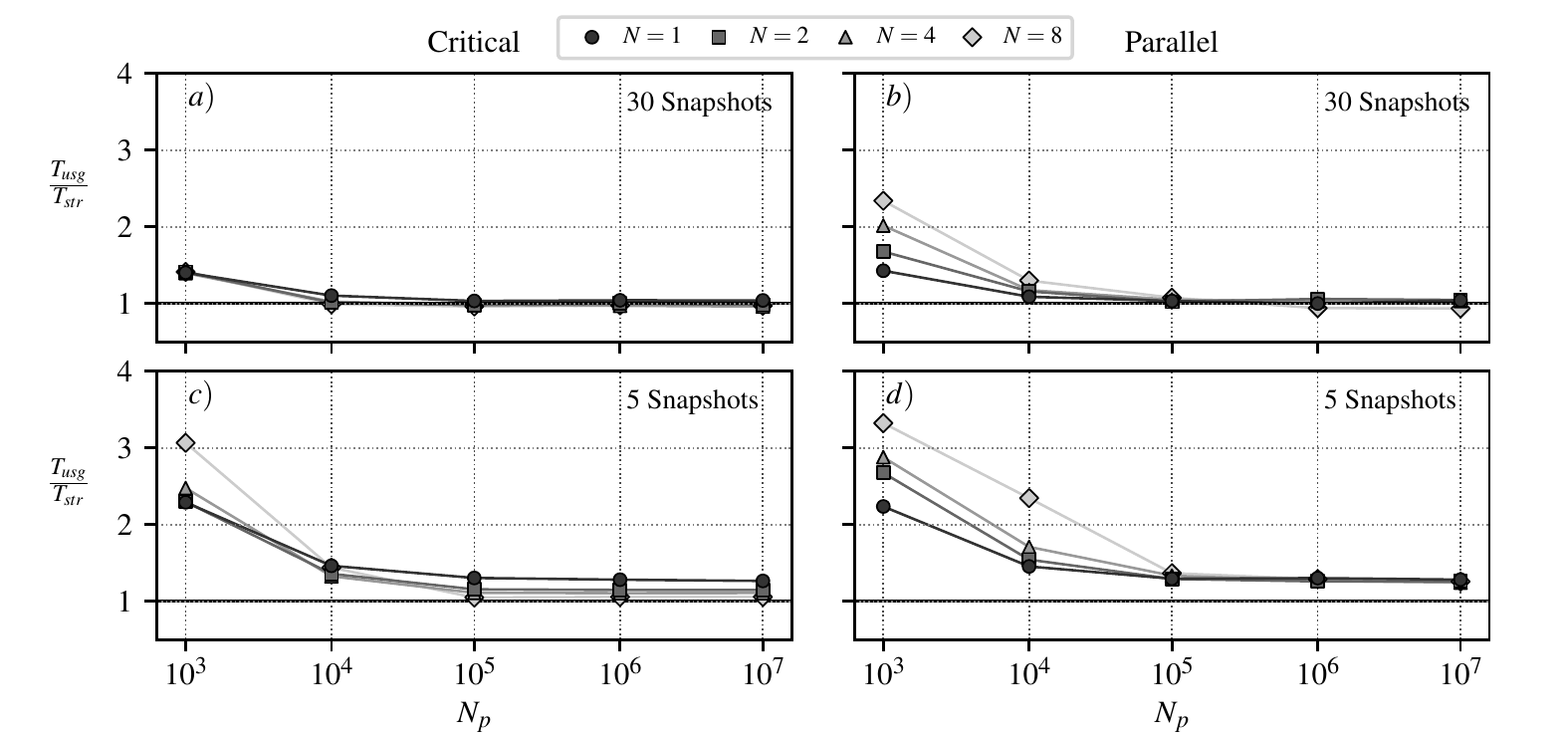}
    \caption{Relative time of simulations with unstructured grid in comparison to structured in test case TC2, for different number of particles and output protocols.}  
    \label{fig:complex:strusg}
\end{figure}

An interesting result is obtained when comparing the computational time of simulations from test case TC2 with unstructured grid versus the regular structured $T_{usg}/T_{str}$ with increasing number of particles. Runs are performed with \texttt{dynamic} scheduling and increasing the number of particles leads to almost the same computational time for the regular and unstructured grids (Fig. \ref{fig:complex:strusg}). These results confirm that above a threshold number of particles, the runtime of the particle tracking model appears to be independent of groundwater flow complexity or the number of cells. It is rather controlled by the total number of particles in concordance with results from both endpoint and timeseries simulations discussed in previous sections.

Overall, results from this test case show that there are scenarios in which better grid model resolutions can be achieved without sacrificing runtime. Notice that when employing the parallel output protocol some difference in runtimes between different flow model grids is preserved for the case with smaller output frequency (TS5). Still, the ratio $T_{usg}/T_{str}$ is close to one, and above $N_p=10^{5}$ remains practically constant for all number of threads. In particle tracking problems with large number of particles, the runtime is controlled by this quantity rather than specific features of the flow model grid, at least for the conditions presented in test case TC2. The extent of parameters up to which these observations remain valid should be further evaluated in future research.

% FIGURE 8

    %% CONCLUSIONS
    \section*{Conclusions}
In this article, the integration of parallel particles processing into the semi-analytical particle tracking program MODPATH has been presented and discussed. Potential for parallel computing was identified from an initial diagnostic of the source code and flow diagram of the program. The parallel particles loop is achieved by integrating the OpenMP library into the Fortran source code, allowing the management of a large number of particles in a desktop computer with efficient runtimes compared to serial processing of particles. Besides OpenMP directives, the implementation required the introduction of an intermediate object class to manage groundwater flow model data that avoids unnecessary replication of arrays when entering the parallel region. 

Two synthetic test cases were used to quantify the improvements in computational times. Results show that speed up factors are limited by the number of particles, meaning that these approach an asymptotic value while increasing total particles for a given number of threads. Endpoint simulations in heterogeneous aquifers of varying degrees of heterogeneity where used to determine an efficient thread scheduling strategy. Analyses showed that \texttt{dynamic} scheduling is convenient for MODPATH in practical groundwater applications, as the time required for displacing each particle naturally adapts to streamline velocities. The influence of the scheduling protocol is not that important in simulations with low spatial variability of groundwater velocity and low number of particles. However, as aquifer heterogeneity and the number of particles increases, the chosen scheduling protocol clearly impacts the overall program performance, most noticeably when employing a number of threads equal to the maximum number of cores of the benchmark system.

For timeseries simulations, three output protocols compatible for parallel computing were discussed. Model runs using thread specific output units where significantly faster than other approaches writing to a single output unit. Moreover, depending on the number of timeseries snapshots, consolidating data from thread specific units into a single file introduces an important computational overhead that ultimately could degrade the advantages provided by parallel processing. Thread specific output files preserves parallelization speed ups, but requires adapting post-processing tools to load data from different files, and if necessary, sorting particles indexes. Output protocols for timeseries were configured to preserve the order of time indexes. The main difference in output files of the parallel implementation in comparison to previous serial models is the sorting of particles indexes. Comparison of simulation runtimes for unstructured and structured grids from second synthetic test case showed that above a certain number of particles, the total simulation time is similar for both grid types. That is, it is possible to obtain a higher spatiotemporal resolution from particle tracking models without significantly sacrificing runtimes.

As MODPATH already provides code infrastructure for managing MODFLOW models, it is a good starting point for the development of particle-based transport models that aim to be integrated with this program. In this regard, it is of relevance to improve the performance of MODPATH base code with parallel particles processing. Parallelization has been implemented minimizing interventions to the current public version of the source code. Future developments may consider further applications of the OpenMP library to other serial stages within the program outside the particles loop, and the integration of additional parallelization methodologies like the Message Passing Interface (MPI) which could be justified for high performance computing architectures simulating several particle groups, each of them with a large number of particles.

    %% ACKNOWLEDGMENTS
    \section*{Acknowledgments}
To Alden Provost, Chris Langevin and Joseph Hughes of the U.S. Geological Survey for discussions regarding MODPATH source code. 

The research leading to these results has received funding from the European Union’s Horizon 2020 research and innovation programme under the Marie Skłodowska ‐ Curie grant agreement no. 814066 ( Managed Aquifer Recharge Solutions Training Network – MARSoluT ).

DFG acknowledges financial support provided by the AGAUR (Agència de Gestió d’Ajuts Universitaris i de Recerca, Generalitat de Catalunya), through project AGAUR-2017-SGR1485.

    %% SUPPORTING
    \section*{Supporting Information}
    Source code for the parallel version of MODPATH: \href{https://github.com/MARSoluT/modpath-omp}{https://github.com/MARSoluT/modpath-omp}. Supporting Information is generally \emph{not} peer reviewed.

    % References
    \printbibliography

    \clearpage
    \pagebreak

    %% CAPTIONS
    \section*{Figure captions}

% FIGURE 1
Figure 1: MODPATH simplified flow chart. Particles loop is parallelized with OpenMP.

% FIGURE 2
Figure 2: OpenMP specification for parallel particles loop in MODPATH. Memory state for all loop variables is explicitly declared as \texttt{shared}, \texttt{private} or \texttt{firstprivate}. Counters are declared with a summation \texttt{reduction} clause.

% FIGURE 3
Figure 3: Synthetic two-dimensional heterogeneous aquifer (TC1). Vertical black line near the origin marks particles injection and white lines show reference streamlines for $\sigma^2_Y=2.5$.

% FIGURE 4
Figure 4: Synthetic three-dimensional layered aquifer (TC2). $a)$ Original structured grid, $b)$ modified unstructured grid. In both panels, scatter points indicate the particles release area.

% FIGURE 5
Figure 5: Performance of endpoint simulations with $N_p=10^7$ particles as a function of aquifer variability and different number of threads $N$. $a)$ Computational time for \texttt{static} (solid, squares) and \texttt{dynamic} (dashed, circles) scheduling with 4 and 8 threads; $b)$ measured ratio $T_{dyn}/T_{sta}$ for all thread configurations.

% FIGURE 6
Figure 6: Speed up for each threading protocol. $a),b)$ as a function of the number of particles with aquifer variance of $\sigma_Y^2=2.5$, $c),d)$ as a function of aquifer variance with number of particles $N_p=10^7$.

% FIGURE 7
Figure 7: Speed up factors for parallel output protocols. Panels $a),c),e)$ (first column) present results for timeseries in test case TC1 with 10 writing stages considering $\sigma^2_Y=2.5$. Panels $b),d),f)$ (second column) display results for test case TC2 with 5 (solid, TS5) and 30 (dashed, TS30) writing stages. Corresponding output protocol is indicated in each panel.

% FIGURE 8 
Figure 8: Relative time of simulations with unstructured grid in comparison to structured in test case TC2, for different number of particles and output protocols.

\end{document}